# Initialization-Free Multistate Memristor: Synergy of Spin-Orbit Torque and Magnetic Fields

Raghvendra Posti [1], Chirag Kalouni [1], Dhananjay K Tiwari [2] and Debangsu Roy[*]

[1]Department of Physics, Indian Institute of Technology Ropar, Rupnagar 140001, India

[2]Silicon Austria Labs GmbH, Sandgasse 34, Graz 8010, Austria

Spin-orbit torque (SOT)-based perpendicularly magnetized memory devices with multistate memory have garnered significant interest due to their applicability in low-power in-memory analog computing. However, current methods are hindered by initialization problems such as prolonged writing duration, and limitations on the number of magnetic states. Consequently, a universal method for achieving multistate in PMA-based stacks remains elusive. Here, we propose a general experimental method for achieving multistate without any initialization step in SOT-driven magnetization switching by integrating an external out-of-plane magnetic field. Motivated by macrospin calculations coupled with micromagnetic simulations, which demonstrate the plausibility of magnetization state changes due to out-of-plane field integration, we experimentally verify multistate behavior in Pt/Co/Pt and W/Pt/Co/AlOx stacks. The occurrence of multistate behavior is attributed to intermediate domain states with Néel domain walls. We achieve repeatable 18 multistate configurations with a minimal reduction in retentivity through energy barrier measurements, paving the way for efficient analog computing.

___

* Corresponding author: debangsu@iitrpr.ac.in

The advent of modern artificial intelligence (AI) has led to an increasing integration of AI across various facets, including industry, research, and everyday life in recent years. However, unlike other technological advancements in modern AI, computing hardware has not progressed at a comparable pace[1]. Operating on the Von Neumann architecture, current computing hardware experiences significant power consumption and latency due to the extensive data transfer between memory and processing units[2]. A recent solution to address the limitations of the Von Neumann architecture is analog computing. This approach utilizes a network of analog memory elements to perform parallel computing directly in the memory[3,4]. While contemporary memory cells operate in a binary configuration, analog memory cells exhibit multistate behavior, enhancing the bit-density per memory cell. For example, an 8-bit sequence could increase its storage capacity by approximately 1500 times by representing five multiple states instead of binary storage[5,6] ($5^8$ vs. $2^8$).

In recent years, spin-orbit torque (SOT)-based perpendicularly magnetized memory devices have emerged as promising candidates for analog computing applications due to their inherent non-volatility, fast operation capabilities, and low power consumption[7,8]. Several reports have demonstrated multistate behavior in SOT devices with perpendicular magnetic anisotropy (PMA). These reports employ different SOT-based mechanisms to illustrate the multistate memory behavior. These mechanisms can be categorized into three distinct types: a) domain nucleation and stabilization of intermediate domain states[9-14], b) magnetization switching of several magnetic layers from a single SOT source and identifying the relative magnetization orientations of magnetic layers[15-19], and c) creating artificial inhomogeneities in devices to control domain motion[20-23]. While the domain state stabilization method can generate a large number of intermediate states, it faces an initialization problem. Before a writing event takes place, the memory cell must reset to an initial state, prolonging the writing duration and consuming additional power. While employing multiple magnetic layers allows for the generation of multiple states without the need for an initialization step, it is hindered by the limited number of magnetic states. For instance, with two magnetic layers, four different states can be produced, necessitating additional material stacking to enhance multisite density[16,18].Additionally, observing multistate behavior by creating inhomogeneities in devices to control domain motion poses experimental complications in terms of rigorous and precise device engineering.

Nevertheless, all the above-mentioned mechanisms are either material stack-specific or device geometry-specific. Thus, there is a lack of a general method that exhibits multistate memory behavior without an initialization step, regardless of material stacking. In this letter, we have endeavoured to establish a universal methodology applicable to any PMA heterostructures, aiming to stabilize multiple states without the need for initialization, and without resorting to complex device or material processing.

In a typical scenario for heavy metal (HM)/ferromagnet (FM) heterostructure with PMA, current-induced magnetization reversal is facilitated through domain nucleation and subsequent domain wall (DW) motion[24,25]. Here, SOT acts as an effective out-of-plane field to the heterostructure plane. It is therefore intuitive to consider that the application of a static magnetic field antiparallel to the SOT field can constrain magnetization saturation, potentially leading to intermediate domain states. Moreover, adjusting the magnitude of the external magnetic field could result in an intermediate domain state, leading to a significant number of multistates. In an attempt to provide a universal experimental method for realizing multistates, we explored the possibility of modifying SOT through the application of an additional out-of-plane static magnetic field ($H_Z$) to the HM/FM plane.

To account for the effect of $H_Z$ on magnetization reversal, we first addressed the magnetization dynamics within the macrospin framework which is described by the Landau-Lifshitz-Gilbert (LLG) equation. After incorporating SOT, the LLG equation is expressed as:

$$\frac{\partial \boldsymbol{m}}{\partial t} = -\gamma\, \boldsymbol{m} \times \boldsymbol{H}_{eff} + \alpha\, \boldsymbol{m} \times \frac{\partial \boldsymbol{m}}{\partial t} + \gamma\, H_{AD}\boldsymbol{m} \times (\boldsymbol{m} \times \boldsymbol{\sigma}) \qquad (1)$$

where $\gamma$ and $\alpha$ are gyromagnetic ratio and Gilbert damping parameter.

The first and second terms of Equation (1) denote the precession around an effective field ($\boldsymbol{H}_{eff}$) and damping of magnetization (**m**). The third term is due to current-induced SOT, where the spin Hall effect (SHE) in HM generates a spin current with a spin polarization direction $\boldsymbol{\sigma}$ (∥ y). Spin currents then exert an anti-damping-like spin-orbit torque (AD-SOT) on FM magnetization (**m**). The AD-SOT is characterized by an effective AD-SOT field $H_{AD}$. Here, FM magnetization **m** ($m_x, m_y, m_z$) is denoted by ($\theta, \varphi$) coordinates (schematic of system is shown in Figure 1 (a)). To achieve deterministic switching, an in-plane symmetry-breaking field ($H_x$) along the current direction (∥x) is applied. Additionally, we incorporated a static field ($H_Z$) out-of-plane to the HM/FM heterostructure to explore its effect on the current generated AD-SOT field ($H_{AD}$). Thus, the net effective field ($\boldsymbol{H}_{eff}$) in Equation (1) is:

$$\boldsymbol{H}_{eff} = H_x\hat{\boldsymbol{x}} + (H_k \cos\theta + H_z)\hat{\boldsymbol{z}} \qquad (2)$$

Here, $H_k$ is the effective perpendicular anisotropy field.

After the current injection the steady state solution of Equation (1) (see supplementary information S1) provides the critical value of SHE induced field ($H_{AD}^C$) as:

$$H_{AD}^C = \frac{1}{\sqrt{2}}(H_z - H_x) + \frac{H_k}{2} \qquad (3)$$

Here, $H_{AD}^C$ is referred as a value of $H_{AD}$ that induces magnetization switching.

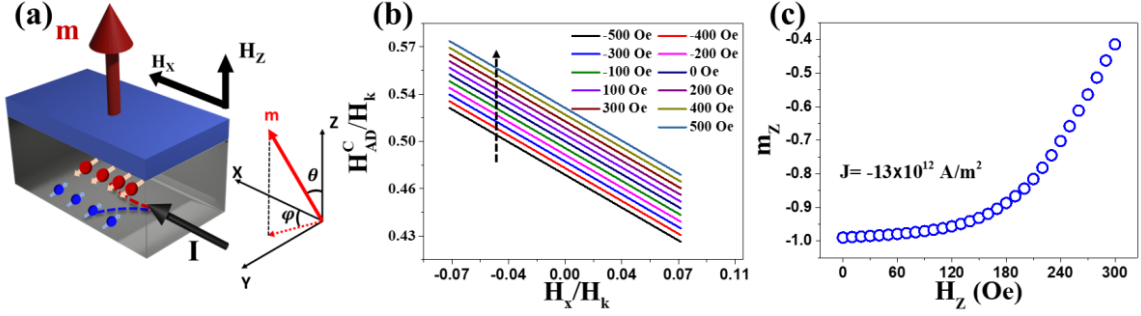

**Figure 1. (a)** Schematic of FM/HM model system with coordinate geometry. In-plane current along x- direction through HM and external fields along x- and z- directions ($H_X$ and $H_Z$, respectively) are applied. **(b)** $H_{AD}^C$ values from Equation (3) as a function of $H_X$ in presence of different $H_Z$ values, where arrow indicates increasing magnitude of $H_Z$ (here, $H_X$ and $H_{AD}$ field values are scaled by anisotropy field $H_k$) **(c)** Results of micromagnetic simulation showing the trend of change in magnetization z-component ($m_z$) after the combined action of $H_Z$ and SOT current pulse.

The macrospin analysis (Equation (3)) suggests that the SOT-induced field required for magnetization switching modifies due to the application of $H_Z$. The graphical representation of Equation (3) is shown in Figure 1(b), where $H_{AD}^C$ been plotted against $H_x$ for different values of $H_Z$. In Figure 1(b), field values were scaled by $H_k$. Here, the value of $H_k$ is taken as a real value of 1.4 T from our previous experimental work[26]. Remarkably, the $H_Z$=0 case replicates the results of previous reports when $H_Z$ was not taken into account[27-30]. It is conceivable that as the critical field for switching increases along with the increase in $H_Z$, a device capable of withstanding a maximum current resulting in a SOT field lower than the critical field for switching may lead to incomplete saturation. In order, to verify the effect of $H_Z$ on the magnetization state, we perform micromagnetic simulations on Pt/Co/Pt material stack (simulation details are in supplementary section S2) and the results are shown in Figure 1(c).Magnetization states were recorded after the concomitant effect of current pulse, symmetry breaking field ($H_X$) and the application of $H_Z$ magnetic field. In Figure 1(c), we observed a trend of decreasing magnetization levels ($m_z$) with the increasing $H_Z$. This is an outcome of Equation (3), which suggests that the application of an out-of-plane field during the current pulse can modify the current-induced effective field and hence can possibly stabilize a different intermediate state (refer to the supplementary section S2 for detailed discussion of micromagnetic results).

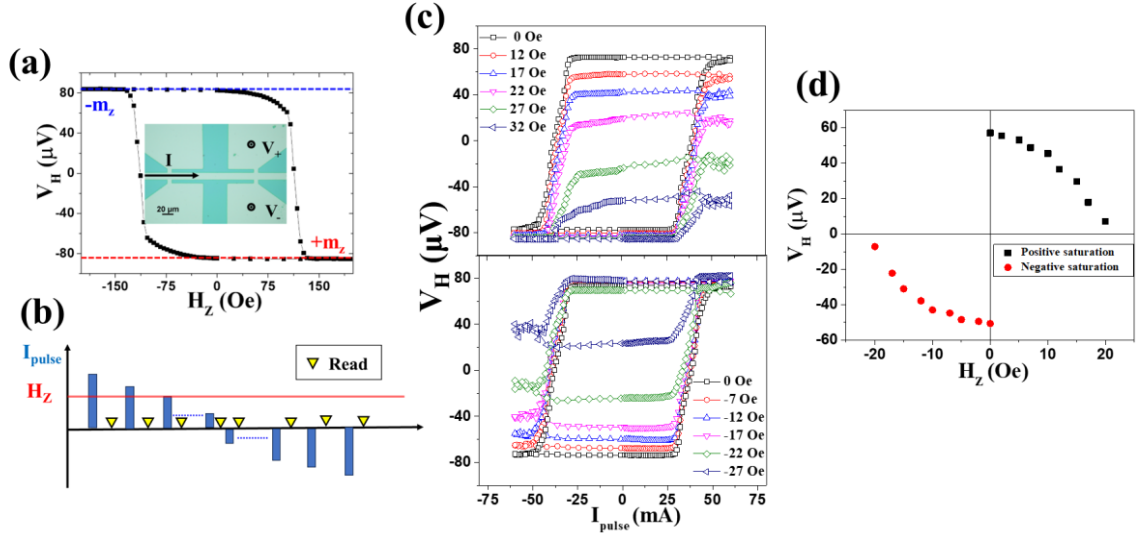

**Figure 2.** **(a)** Anomalous Hall effect (AHE) voltage ($V_H$) vs. $H_Z$. Inset: Optical image of Hall bar device with measurement geometry. **(b)** Procedure for measuring magnetization switching in the presence of $H_Z$ field, here, after each current pulse magnetization is read through AHE voltage ($V_H$). **(c)** SOT switching in presence of positive (top) and negative (bottom) $H_Z$ fields showing multistate behavior. **(c)** Trend of reduction for magnetization states in SOT switching in presence of different $H_Z$ magnitude and polarity.

For the experimental verification, we have selected, Si/SiO$_2$/Ta(3 nm)/Pt(3 nm)/Co(0.6 nm)/Pt(6 nm) stack. The reason for selecting Pt/Co/Pt system for this study is twofold: Pt/Co/Pt exhibit a strong PMA (such that $H_x$, $H_Z$ << $H_K$, see supplementary information S1) which is a necessary requirement for realizing dense memory cells[31]. Additionally, Pt/Co/Pt is one of the earliest model systems for PMA studies[26,32-34] thereby expanding the potential applicability of the current results. The deposition and fabrication processes were previously outlined in our earlier work[26]. Figure 2(a) (inset) depicts a schematic overview of the 12 μm wide Hall bar device along with the measurement geometry used in all room temperature transport investigations. The anomalous Hall effect (AHE) measurements validate the PMA in the Pt/Co/Pt stack. Additionally, the lower (higher) magnitude of the Hall voltage distinguishes the +m$_z$ (-m$_z$) state. The schematic of the SOT induced magnetization switching scheme in presence of Hz is shown in Figure 2(b). Here, 100 μs current pulses ($I_{pulse}$) are applied in presence of a $H_X$ and subsequently the magnetization orientation is read by conventional AHE ($V_H$) measurements utilizing a small reading current of 1 mA. During an individual current-induced switching measurement through SOT, a fixed Hz is maintained. In this manuscript, we varied the magnitude and polarity of Hz during different current sweeps, as illustrated in Figure 2(c). At each Hz value, $V_H$ vs. $I_{pulse}$ hysteresis was recorded. Remarkably, the introduction of an additional field Hz results in the emergence of multistate analog switching behavior (as illustrated in Figure 2 (c)). Here, $H_x$ is ~900 Oe, and Hz of varying magnitude and polarities are applied. The top (bottom) part of the Figure 2(c) depicts the multistate behaviour in the studied stack in presence of positive (negative) Hz. Note that the magnitude of $H_Z$ is always less than coercive field ($H_c$ ~ 110 Oe) of the stack. Here, the application of a small + $H_Z$(–$H_Z$) fixes the + $m_z$ (–$m_z$) states whereas shrink of – $m_z$ (+$m_z$) state is observed. Notably, the changes in magnetization states are more prominent with increasing magnitude of Hz. Figure 2(d) illustrates the trend of magnetization state reduction with Hz. Importantly, the experimentally observed trend of magnetization reduction resembles the trend obtained from micromagnetic simulation. This similarity reflects that the reduction in the magnitude of the magnetization state is an outcome of SOT field modification due to Hz. Since the acquired AHE signal ($V_H$) is proportional to the weighted average of the z-component of magnetization ($m_z$) in the heterostructure, the reduction in $V_H$ values suggest a domain state formation which reduces the $m_z$ magnetization state magnitude[35,36]. Therefore, the reduced magnetization values of +$m_z$ (-$m_z$) for -Hz (+Hz) are possibly the attributes of intermediate domain states.

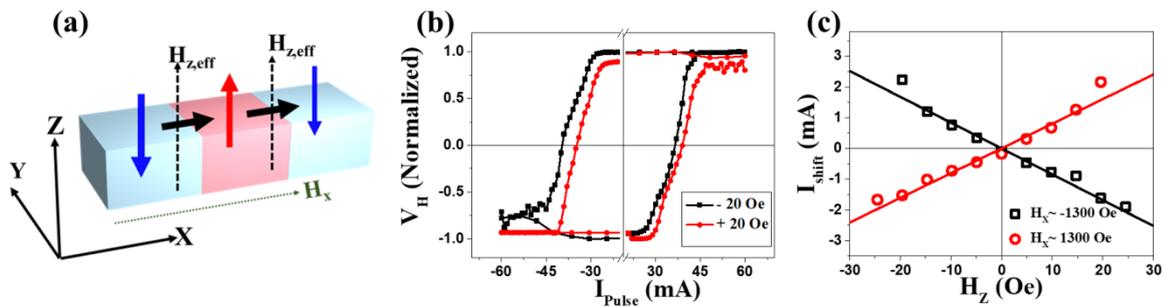

**Figure 3.** **(a)** Out of plane effective SOT field ($H_{z,eff}$) acting on chirality broken Néel DW. **(b)** SOT loop shift as a consequence of ±Hz. Here, a symmetry breaking field $H_x$~1300 Oe was also applied along the current direction. **(c)** Linear trend of SOT induced hysteresis loop shift as a function of Hz. Linear plots with opposite slope attributes to the opposite polarity of $H_x$.

In the preceding section, we speculated the stabilization of the intermediate domain state which indicates that the magnetization reversal mechanism could occur due to domain wall nucleation and its subsequent propagation. The field-induced magnetization reversal indeed follows domain wall nucleation and subsequent propagation in comparison to the coherent rotation of the magnetization reversal as revealed by the angle-dependent coercive field measurement and its subsequent fitting using the Kondorsky model[24,37,38] (see supplementary information S4). In SOT-induced switching, the magnetization reversal occurs after the introduction of $H_X$ along the current direction[39]. This field stabilizes Néel DWs with a broken chiral symmetry where the in-plane projection of DW magnetization ($m_x$) align along $H_X$ field direction[24,25,40] (shown in Figure 3(a)). On these achiral DW, AD-SOT acts as an effective field along z-direction ($H_{z,eff}$), defined by $H_{z,eff} \propto (m \times \sigma) \sim m_x \times \sigma$. Therefore, AD-SOT leads to an asymmetric motion of the chirality broken Néel DW and eventually leads to the deterministic magnetization reversal (Figure 3(a)). A charge current applied to the heterostructure generates $H_{z,eff}$ whose magnitude and direction depends on magnitude of current (I) and DW profile as[25,41,42]:

$$H_{z,eff} = \frac{\pi}{2} \chi_{AD} \cos\varphi_{DW} I_\square \qquad (4)$$

Where, $\chi_{AD} \left(= \frac{H_{AD}}{I_\square} = \frac{\hbar \theta_{SH}}{2e\mu_0 M_s t}\right)$ is AD-SOT efficiency, and $\theta_{SH}$, Ms, t corresponds to spin Hall angle of HM layer, saturation magnetization and thickness of FM layer, respectively, and $\varphi_{DW}$ corresponds to the angle between current direction and DW magnetization.

Thus, when the symmetry-breaking field $H_X$ stabilizes an achiral configuration of Néel DW, according to Equation (4), the SOT effective field scales with the current. Equation (4) also suggests that an additional field along the z-direction will lead to an offset along the current axis of the SOT switching ($V_H$ vs. $I_{pulse}$) hysteresis loop. Therefore, in the presence of achiral Néel DW, $H_Z$ field will introduce a loop shift in SOT switching hysteresis. This shift will depend on the magnitude and polarity of the $H_Z$ field. Following this, we conducted $V_H$ vs. $I_{pulse}$ hysteresis measurement in the presence of a constant $H_Z$ field. The measurement (Figure 3(b)) reveals a shift in the current assisted switching hysteresis for a constant $H_Z$ field. Furthermore, we have observed an opposing loop shift for the different polarity of the $H_Z$ field ($H_Z$=+20 Oe, and -20 Oe) at a fixed value of $H_X$~1300 Oe, corroborating our macrospin analysis (see supplementary information S1). This loop shift in the SOT current ($I_{pulse}$) axis exhibits (Figure 3(c)) a linear trend for different $H_Z$ field. In addition, the polarity of slope changes depending on the polarity of the applied $H_X$ (whether $H_x$~ +1300 Oe, or -1300 Oe) validating the presence of Néel DW in our stack. The application of a DC field ±$H_Z$ (± sign defines the direction of $H_Z$ along ±z direction) adds or subtract with the SOT effective field $H_{z,eff}$ depending upon the current polarity. For +$I_{Pulse}$, $H_{z,eff}$ is along +z and hence net effective field of SOT and applied DC sums up and favors +$m_z$ saturation state. However, as the polarity of current pulse reverses (-$I_{Pulse}$) the SOT field and applied DC field oppose each other. This leads to a scenario where the resultant field along -z direction does not lead to a complete saturation along -$m_z$ (i.e., $m_z$<-1) and a domain state is stabilized. Thus, the reduction of -$m_z$ in this case depends on the magnitude of $H_Z$.

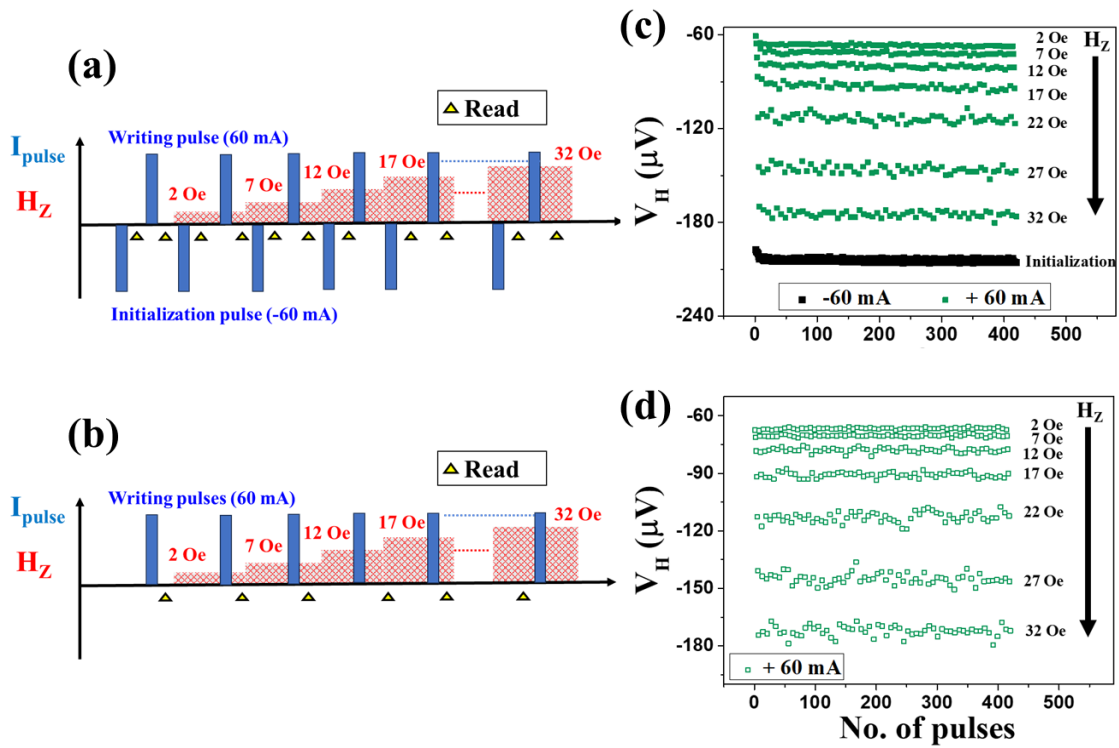

**Figure 4**. Multistate SOT switching procedure **(a)** with initialization step, **(b)** without initialization step. For initialization a -60 mA current was applied and for both cases writing current pulse was + 60 mA. **(c)** Each data points corresponding to the multistates (red points) are achieved after initialization step (black points) **(d)** Occurrence of multistate data without initialization step. ($H_Z$ field values are mentioned along with each state).

Thus far, we have demonstrated that the multistate memory behavior can be stabilized simply by including a static field. We have observed 9 distinct states for each polarity of $H_Z$, totalling 18 states. This number can be further increased by fine-tuning the out-of-plane static field. However, for practical implementation of a multistate memory cell in computing applications, eliminating the initialization step is essential. Our micromagnetic simulations and subsequent experimental results indicate that magnetic state variation is influenced by both SOT and $H_Z$. We then compare the multistate behavior in our PMA stack with and without an initialization step and the writing mechanism pertaining to it is depicted in Figures 4(a) and 4(b), respectively. In both cases, a +60 mA current pulse is applied in the presence of different $H_Z$ fields as part of the writing sequence. Additionally, when the initialization step is included (Figure 4(a)), a -60mA current pulse is applied before each writing pulse (+60 mA) to establish an initial state. After each current pulse, conventional AHE measurements are used to read the magnetic state in the device, and this procedure is repeated over 400 pulses to check its repeatability. Figures 4(c) and 4(d) shows the magnitude of the multistate for with and without the initialization step at different $H_Z$. It is evident that the magnitude of the AHE voltage in Figure 4(c) and (d) remains consistent regardless of whether the initialization step is included or not, rendering the initialization step redundant (see supplementary information S5). Furthermore, based on our theoretical perspective, the occurrence of the multistate behavior, resulting from the application of a static field, should hold true for all PMA heterostructures. To validate this assertion, we conducted SOT switching experiments using the W/Pt/Co/AlOx stack with PMA, revealing a multistate signature exhibiting a comparable non-linear trend of reduction in magnetizationstate (see supplementary information S8).

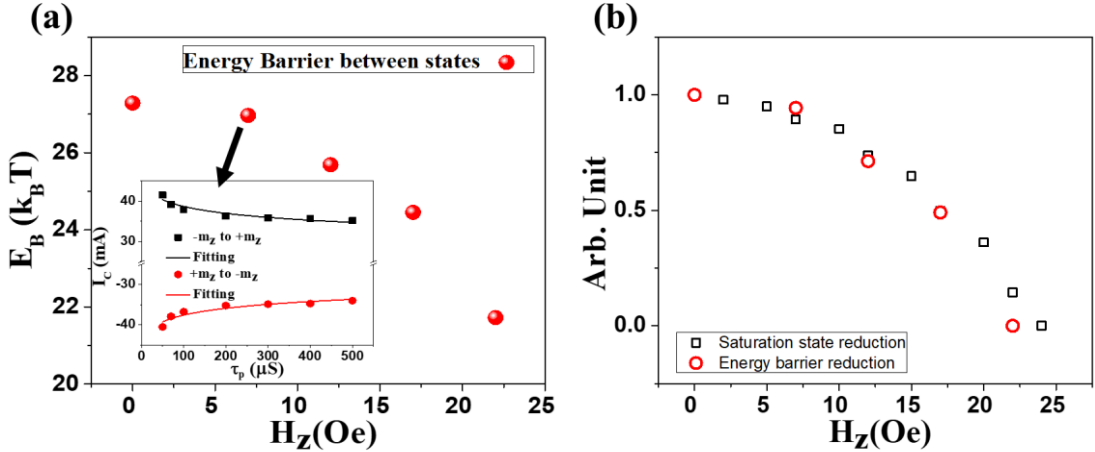

**Figure 5.(a)** Energy barrier obtained from the Arrhenius fitting (Equation (5)) at different $H_Z$ field values (Inset: Representative fitting of Equation (5) when $H_Z$= 7 Oe). **(b)** Comparison of energy barrier (red circle) reduction and reduction of magnetizationstates (black square) as a function of $H_Z$.

Upon verifying the initialization-free multistate behavior, we explored the energy landscape associated with magnetization switching. Analyzing the energy barrier between two stable magnetization state provides insight into the retention properties and switching mechanism. Here, we measured SOT induced switching hysteresis at each $H_Z$ by applying current with different pulse widths. The critical current density in our experiments exhibit dependency on the applied current pulse duration, under a fixed $H_Z$ field. This dependency implies that the SOT switching mechanism is thermally activated, consistent to previous reports[24]. Therefore, experimentally observed critical switching currents ($I_C$) at different pulse width ($\tau_p$) can be effectively analyzed using Arrhenius fitting[43]:

$$I_C = I_{C_0}\left[1 - \frac{k_B T}{E_b} \ln\left(\frac{\tau_p}{\tau_0}\right)\right] \qquad (5)$$

Here, $I_{C_0}$ is critical current value at temperature T=0 K, $\tau_0$ is the attempt switching time, considered as 1 nS. For a fix value of $H_Z$=7 Oe the fitting of Equation (5) is shown in the inset of Figure 5(a).

Further, the energy barrier ($E_b$) between magnetization states as a function of $H_Z$ field is shown in Figure 5 (a). Here, the thermal energy barrier, ~27 $K_B T$ (in the absence of $H_Z$), reduces with the increasing field values. Although the thermal barrier of ~27 $K_B T$ is lesser than the industrial requirements ($E_b$ > 45 $k_B T$ for the data retention period of 10 years),nonetheless, it is comparable to results from recent reports involving Hall bar devices[44-47]. Since this multistate behavior of SOT switching with $H_Z$field is a generalizedexperimental method for all PMA stacks, selecting an appropriate stack would enable the attainment of the desired data retention time. Further, the energy barrier is reduced by ~18% from $H_Z$=0 Oe to 23 Oe. Since the energy barrier between stable configurations is a measure of memory retention or stability, therefore,the retention time is reduced minimally after a large number of states. Furthermore, upon comparing the normalized data of the energy barrier and the reduced magnetizationstate (as depicted in Figure 5 (b)), we observed significant similarities in their trends. This finding is logical as the saturation state for the $H_Z$ = 0 Oe case corresponds to the complete saturation of the ferromagnetic state where as for $H_Z$ ≠ 0 Oe, intermediate domain states form, which possess lower energy levels than the complete saturation state.

In conclusion, we utilized the combined symmetry of SOT and a static field to stabilize multiple states within PMA heterostructures. Our investigation confirmed that these multistates are characterized by both the magnitudes and polarity of the static field. These multistates represent distinct domain states resulting from SOT manipulation following the application of a static field ($H_Z$). Our results suggest that this phenomenon arises from the concurrent influence of both SOT and the applied field. Importantly, our experimental measurements effectively address the initialization problem in multistate SOT devices by demonstrating initialization-free multistate memory. These findings pave the way for the development of high-bit-density memory cells for analog computing applications and its subsequent integration with the existing device technology.

*Acknowledgements*—DR acknowledge the financial support from the Department of Atomic Energy (DAE) under project no. 58/20/10/2020-BRNS/37125 & Science and Engineering Research Board (SERB) under project no. CRG/2020/005306.

## DATA AVAILABILITY

The data that support the findings of this study are available from the corresponding author upon reasonable request.

# References

[1]     Big data needs a hardware revolution, Nature **554**, 145 (2018).
[2]     J. Backus, Can programming be liberated from the von Neumann style? a functional style and its algebra of programs, **21**, 613 (1978).
[3]     M. Di Ventra and Y. V. Pershin, The parallel approach, Nature Physics **9**, 200 (2013).
[4]     J. Grollier, D. Querlioz, K. Y. Camsari, K. Everschor-Sitte, S. Fukami, and M. D. Stiles, Neuromorphic spintronics, Nature Electronics **3**, 360 (2020).
[5]     Q. Cao, W. Lü, X. R. Wang, X. Guan, L. Wang, S. Yan, T. Wu, and X. Wang, Nonvolatile Multistates Memories for High-Density Data Storage, ACS Applied Materials & Interfaces **12**, 42449 (2020).
[6]     R. Morales, M. Kovylina, I. K. Schuller, A. Labarta, and X. Batlle, Antiferromagnetic/ferromagnetic nanostructures for multidigit storage units, Applied Physics Letters **104**, 032401 (2014).
[7]     B. Dieny and M. Chshiev, Perpendicular magnetic anisotropy at transition metal/oxide interfaces and applications, Reviews of Modern Physics **89**, 025008 (2017).
[8]     A. Manchon, J. Železný, I. M. Miron, T. Jungwirth, J. Sinova, A. Thiaville, K. Garello, and P. Gambardella, Current-induced spin-orbit torques in ferromagnetic and antiferromagnetic systems, Reviews of Modern Physics **91**, 035004 (2019).
[9]     A. Kurenkov, C. Zhang, S. DuttaGupta, S. Fukami, and H. Ohno, Device-size dependence of field-free spin-orbit torque induced magnetization switching in antiferromagnet/ferromagnet structures, Applied Physics Letters **110**, 092410 (2017).
[10]    J. Zhou *et al.*, Spin–Orbit Torque-Induced Domain Nucleation for Neuromorphic Computing, Advanced Materials **33**, 2103672 (2021).
[11]    K. Olejník *et al.*, Antiferromagnetic CuMnAs multi-level memory cell with microelectronic compatibility, Nature Communications **8**, 15434 (2017).
[12]    Y. Yang, H. Xie, Y. Xu, Z. Luo, and Y. Wu, Multistate Magnetization Switching Driven by Spin Current From a Ferromagnetic Layer, Physical Review Applied **13**, 034072 (2020).
[13]    K. Dong, Z. Guo, Y. Jiao, R. Li, C. Sun, Y. Tao, S. Zhang, J. Hong, and L. You, Field-Free Current-Induced Switching of L10-$\mathrm{Fe}\mathrm{Pt}$ Using Interlayer Exchange Coupling for Neuromorphic Computing, Physical Review Applied **19**, 024034 (2023).
[14]    S. Fukami, C. Zhang, S. DuttaGupta, A. Kurenkov, and H. Ohno, Magnetization switching by spin–orbit torque in an antiferromagnet–ferromagnet bilayer system, Nature Materials **15**, 535 (2016).
[15]    H. Fan *et al.*, Field-free switching and high spin–orbit torque efficiency in Co/Ir/CoFeB synthetic antiferromagnets deposited on miscut Al2O3 substrates, Applied Physics Letters **122**, 262404 (2023).
[16]    C. O. Avci, M. Mann, A. J. Tan, P. Gambardella, and G. S. D. Beach, A multi-state memory device based on the unidirectional spin Hall magnetoresistance, Applied Physics Letters **110**, 203506 (2017).
[17]    Y. Song, X. Zhao, W. Liu, L. Liu, S. Li, and Z. Zhang, Spin–orbit torque driven four-state switching in splicing structure, Applied Physics Letters **117**, 232408 (2020).
[18]    Z. Zheng *et al.*, Enhanced Spin-Orbit Torque and Multilevel Current-Induced Switching in $\mathrm{W}/\mathrm{Co}\text{\ensuremath{-}}\mathrm{Tb}/\mathrm{Pt}$ Heterostructure, Physical Review Applied **12**, 044032 (2019).
[19]    Y. Sheng, Y. C. Li, X. Q. Ma, and K. Y. Wang, Current-induced four-state magnetization switching by spin-orbit torques in perpendicular ferromagnetic trilayers, Applied Physics Letters **113**, 112406 (2018).
[20]    J. Cai, B. Fang, C. Wang, and Z. Zeng, Multilevel storage device based on domain-wall motion in a magnetic tunnel junction, Applied Physics Letters **111**, 182410 (2017).
[21]    J. Kurian *et al.*, Deterministic multi-level spin orbit torque switching using focused He+ ion beam irradiation, Applied Physics Letters **122**, 032402 (2023).

[22]  S. A. Siddiqui, S. Dutta, A. Tang, L. Liu, C. A. Ross, and M. A. Baldo, Magnetic Domain Wall Based Synaptic and Activation Function Generator for Neuromorphic Accelerators, Nano Letters **20**, 1033 (2020).
[23]  H. Mohammed, S. A. Risi, T. L. Jin, J. Kosel, S. N. Piramanayagam, and R. Sbiaa, Controlled spin-torque driven domain wall motion using staggered magnetic wires, Applied Physics Letters **116**, 032402 (2020).
[24]  O. J. Lee, L. Q. Liu, C. F. Pai, Y. Li, H. W. Tseng, P. G. Gowtham, J. P. Park, D. C. Ralph, and R. A. Buhrman, Central role of domain wall depinning for perpendicular magnetization switching driven by spin torque from the spin Hall effect, Physical Review B **89**, 024418 (2014).
[25]  C.-F. Pai, M. Mann, A. J. Tan, and G. S. D. Beach, Determination of spin torque efficiencies in heterostructures with perpendicular magnetic anisotropy, Physical Review B **93**, 144409 (2016).
[26]  R. Posti, A. Kumar, D. Tiwari, and D. Roy, Emergence of considerable thermoelectric effect due to the addition of an underlayer in Pt/Co/Pt stack and its application in detecting field free magnetization switching, Applied Physics Letters **121**, 223502 (2022).
[27]  D. Zhu and W. Zhao, Threshold Current Density for Perpendicular Magnetization Switching Through Spin-Orbit Torque, Physical Review Applied **13**, 044078 (2020).
[28]  T. Taniguchi, Theoretical condition for switching the magnetization in a perpendicularly magnetized ferromagnet via the spin Hall effect, Physical Review B **100**, 174419 (2019).
[29]  T. Taniguchi, S. Mitani, and M. Hayashi, Critical current destabilizing perpendicular magnetization by the spin Hall effect, Physical Review B **92**, 024428 (2015).
[30]  K.-S. Lee, S.-W. Lee, B.-C. Min, and K.-J. Lee, Threshold current for switching of a perpendicular magnetic layer induced by spin Hall effect, Applied Physics Letters **102**, 112410 (2013).
[31]  S. Mangin, D. Ravelosona, J. A. Katine, M. J. Carey, B. D. Terris, and E. E. Fullerton, Current-induced magnetization reversal in nanopillars with perpendicular anisotropy, Nature Materials **5**, 210 (2006).
[32]  A. Kobs, S. Heße, W. Kreuzpaintner, G. Winkler, D. Lott, P. Weinberger, A. Schreyer, and H. P. Oepen, Anisotropic Interface Magnetoresistance in $\mathrm{Pt}/\mathrm{Co}/\mathrm{Pt}$ Sandwiches, Physical Review Letters **106**, 217207 (2011).
[33]  P. J. Metaxas, J. P. Jamet, A. Mougin, M. Cormier, J. Ferré, V. Baltz, B. Rodmacq, B. Dieny, and R. L. Stamps, Creep and Flow Regimes of Magnetic Domain-Wall Motion in Ultrathin $\mathrm{Pt}/\mathrm{Co}/\mathrm{Pt}$ Films with Perpendicular Anisotropy, Physical Review Letters **99**, 217208 (2007).
[34]  R. Lavrijsen, D. M. F. Hartmann, A. van den Brink, Y. Yin, B. Barcones, R. A. Duine, M. A. Verheijen, H. J. M. Swagten, and B. Koopmans, Asymmetric magnetic bubble expansion under in-plane field in Pt/Co/Pt: Effect of interface engineering, Physical Review B **91**, 104414 (2015).
[35]  N. Nagaosa, J. Sinova, S. Onoda, A. H. MacDonald, and N. P. Ong, Anomalous Hall effect, Reviews of Modern Physics **82**, 1539 (2010).
[36]  R. Posti, A. Kumar, M. Baghoria, B. Prakash, D. Tiwari, and D. Roy, Odd symmetry planar Hall effect: A method of detecting current-induced in-plane magnetization switching, Applied Physics Letters **122**, 152405 (2023).
[37]  S. Kim, P.-H. Jang, D.-H. Kim, M. Ishibashi, T. Taniguchi, T. Moriyama, K.-J. Kim, K.-J. Lee, and T. Ono, Magnetic droplet nucleation with a homochiral N\'eel domain wall, Physical Review B **95**, 220402 (2017).
[38]  E. J. J. P. Kondorsky, On hysteresis in ferromagnetics, **2**, 161 (1940).
[39]  L. Liu, O. J. Lee, T. J. Gudmundsen, D. C. Ralph, and R. A. Buhrman, Current-Induced Switching of Perpendicularly Magnetized Magnetic Layers Using Spin Torque from the Spin Hall Effect, Physical Review Letters **109**, 096602 (2012).
[40]  P. P. J. Haazen, E. Murè, J. H. Franken, R. Lavrijsen, H. J. M. Swagten, and B. Koopmans, Domain wall depinning governed by the spin Hall effect, Nature Materials **12**, 299 (2013).
[41]  A. Thiaville, S. Rohart, É. Jué, V. Cros, and A. Fert, Dynamics of Dzyaloshinskii domain walls in ultrathin magnetic films, Europhysics Letters **100**, 57002 (2012).
[42]  T. Dohi, S. Fukami, and H. Ohno, Influence of domain wall anisotropy on the current-induced hysteresis loop shift for quantification of the Dzyaloshinskii-Moriya interaction, Physical Review B **103**, 214450 (2021).
[43]  R. H. Koch, J. A. Katine, and J. Z. Sun, Time-Resolved Reversal of Spin-Transfer Switching in a Nanomagnet, Physical Review Letters **92**, 088302 (2004).

[44]	T. Fan, N. H. D. Khang, T. Shirokura, H. H. Huy, and P. N. Hai, Low power spin–orbit torque switching in sputtered BiSb topological insulator/perpendicularly magnetized CoPt/MgO multilayers on oxidized Si substrate, Applied Physics Letters **119**, 082403 (2021).
[45]	W.-B. Liao, T.-Y. Chen, Y.-C. Hsiao, and C.-F. Pai, Pulse-width and temperature dependence of memristive spin–orbit torque switching, Applied Physics Letters **117**, 182402 (2020).
[46]	Y. Saito, S. Ikeda, and T. Endoh, Enhancement of current to spin-current conversion and spin torque efficiencies in a synthetic antiferromagnetic layer based on a Pt/Ir/Pt spacer layer, Physical Review B **105**, 054421 (2022).
[47]	Y.-H. Huang, J.-H. Han, W.-B. Liao, C.-Y. Hu, Y.-T. Liu, and C.-F. Pai, Tailoring Interlayer Chiral Exchange by Azimuthal Symmetry Engineering, Nano Letters **24**, 649 (2024).

# Supporting Information

## Initialization-Free Multistate Memristor: Synergy of Spin-Orbit Torque and Magnetic Fields

Raghvendra Posti [1], Chirag Kalouni [1], Dhananjay K Tiwari [2] and Debangsu Roy[*]

[1]Department of Physics, Indian Institute of Technology Ropar, Rupnagar 140001, India

[2]Silicon Austria Labs GmbH, Sandgasse 34, Graz 8010, Austria

## Table of contents:

___

* Corresponding author: debangsu@iitrpr.ac.in

# S1. Effect of external field (H$_Z$) on SOT:

We explored the possibility of altering the anti-damping torque (AD-SOT) by an application of an external field.

The time evolution of magnetization in the presence of SOTs is governed by the following Landau-Lifshitz Gilbert (LLG) equation:

$$\frac{\partial \boldsymbol{m}}{\partial t} = -\gamma\, \boldsymbol{m} \times \boldsymbol{H}_{eff} + \alpha\, \boldsymbol{m} \times \frac{\partial \boldsymbol{m}}{\partial t} + \gamma\, H_{AD}\, \boldsymbol{m} \times (\boldsymbol{m} \times \boldsymbol{\sigma}) \qquad \text{S(1)}$$

where $\gamma$ and $\alpha$ are gyromagnetic ratio and Gilbert damping parameter, $\boldsymbol{\sigma}$ (∥ y) is spin current polarization, and H$_{AD}$ is an anti-damping effective field produced by SHE.

A symmetry-breaking magnetic field along the current direction (H$_x$) is necessary to induce the SOT switching. Additionally, we apply an external field (H$_Z$) oriented perpendicular to the ferromagnetic (FM)/ Heavy metal (HM) heterostructure plane. The effective field $\boldsymbol{H}_{eff}$ in Eq. S(1), other than H$_x$ and H$_Z$, has a perpendicular anisotropy field contribution (H$_k$). Therefore, the $\boldsymbol{H}_{eff}$ is written as:

$$\boldsymbol{H}_{eff} = (H_x,\ 0,\ H_k \cos\theta + H_z) \qquad \text{S(2)a}$$

The magnetization direction in spherical polar coordinates is defined as:

$$\boldsymbol{m} = (\sin\theta\cos\varphi,\ \sin\theta\sin\varphi,\ \cos\theta) \qquad \text{S(2)b}$$

A charge current in-plane (∥ x) to the HM layer generates a spin current along the z-direction with its spin-polarized along the y-direction. Therefore, the coordinate vector for the polarization of spin current is:

$$\boldsymbol{\sigma} = (0,\ 1,\ 0) \qquad \text{S(2)c}$$

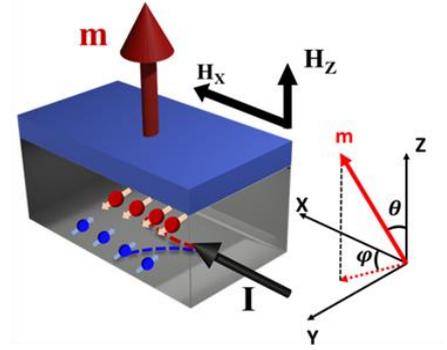

**Fig. S1 (a)**: Schematic of FM/HM system with current, magnetization, and spin polarization directions along with the coordinate geometry.

An illustration of the above-mentioned system is shown in Fig. S1(a).

The equilibrium state of magnetization after the current application can be found form the stationary state solution of Eq. S(1), i.e., $\frac{dm}{dt} = 0$.

$$\Rightarrow \quad -\boldsymbol{m} \times \boldsymbol{H}_{eff} + H_{AD}\, \boldsymbol{m} \times (\boldsymbol{m} \times \boldsymbol{\sigma}) = 0 \qquad \text{S(3)}$$

Using Eq. S(2) in Eq. S(3), we obtain the following equations

$$sin\theta\, sin\varphi(-H_z + H_{AD} sin\theta\, cos\varphi - H_K cos\theta) = 0 \qquad \text{S(4)}$$

$$-H_{AD} cos^2\theta + sin\theta\, cos\varphi(H_z - H_{AD} sin\theta\, cos\varphi) + H_K\, cos\theta\, sin\theta\, cos\varphi - H_x\, cos\theta = 0 \qquad \text{S(5)}$$

$$sin\theta\, sin\varphi(H_{AD}\, cos\theta + H_k) = 0 \qquad \text{S(6)}$$

A critical value of $H_{AD}$ (i.e., $H_{AD}^C$) is responsible for SOT-induced switching of $m_z$. During SOT-induced magnetization switching[1] $m_y \sim 0$, hence, $\varphi = 0$.

Further, $\varphi = 0$ makes Eq. S(4) & S(6) self-evident and Eq. S(5) became:

$$-H_{AD}^C cos^2\theta + sin\theta\, (H_z - H_{AD}^C sin\theta) + H_K\, cos\theta\, sin\theta - H_x\, cos\theta = 0$$

Thus, $H_{AD}^C$ is determined as:

$$h_{AD}^C = h_z\, sin\theta + cos\theta(cos\theta - h_x) \qquad \text{S(7)}$$

Here, $h_{AD}^C$, $h_z$, and $h_x$ are $H_{AD}^C$, $H_Z$, and $H_x$ fields, respectively which are scaled by $H_k$.

i.e., $h_{AD}^C = H_{AD}^C/H_k$, $h_z = H_z/H_k$, and $h_x = H_x/H_k$

To find the values of $sin\theta$ and $cos\theta$ we find the local minima and maxima of Eq. S(7)[1,2]

Which gives

$$4sin^4\theta + 4\, h_x\, sin^3\theta + \left(h_x^2 + h_z^2 - 4\right) sin^2\theta + 2\, h_x\, sin\theta + \left(1 - h_z^2\right) = 0 \qquad \text{S(8)}$$

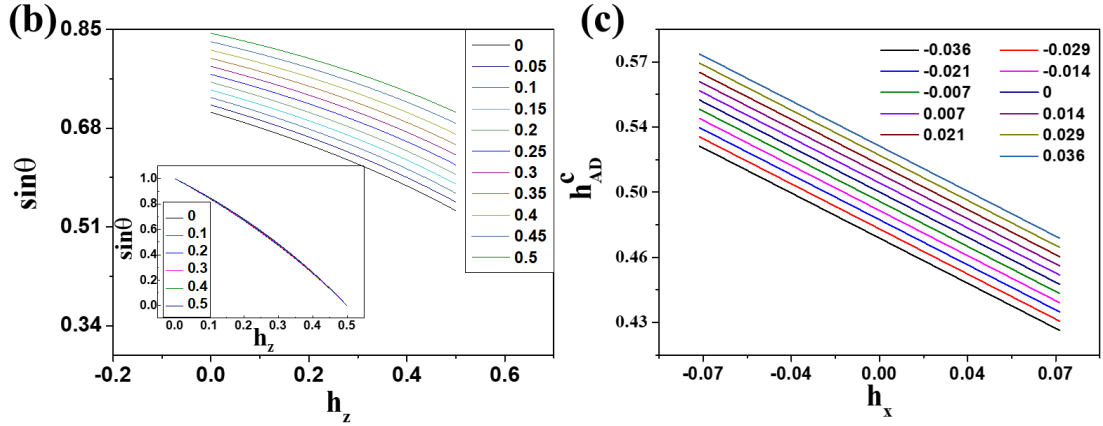

**Fig. S1 (b)** Roots of Eq. S(8) as a function of $h_z$ (=$H_Z/H_k$). Here, various line plots represent the $sin\theta$ vs. $h_z$ plot for a fixed value of $h_x$ (=$H_x/H_k$) (numeric values of $h_x$ are written to the right side of the plot). Inset: normalized $sin\theta$ vs. $h_z$ plot for different values of $h_x$. **(c)** Graphical representation of Eq. S(11) as $h_{AD}^C$ vs. $h_x$ for different values of $h_z$ (values written inside the graph correspond to different $h_z$ values).

The graphical solution of $sin\theta$ was obtained from Eq. S(8). For the SOT switching scenario, $\theta \to 90°$ and Figure S1(b) illustrates these solutions. In Fig. S1(b), the roots of Eq. S(8) are plotted as a function of $h_z$ with varying $h_x$.

Previous research extensively[1-4] explored the general scenario of SOT-induced magnetization switching (where $H_x \neq 0$ and $H_Z = 0$). In those studies, without considering the influence of $H_Z$, the solution for $sin\theta$ takes the following form:

$$sin\theta|_{h_z=0} = \frac{1}{4}\left(h_x + \sqrt{h_x^2 + 8}\right) \quad\quad\quad S(9)$$

We obtained the same solution of Eq. S(8) for the $h_z = 0$ case. In Fig. S1(b), the initial points corresponding to $h_z=0$ for different $h_x$ values follow the solution of $sin\theta|_{h_z=0}$. Upon normalizing the solution of $sin\theta$ as a function of $h_z$ for different $h_x$ values, the graphical trend coincides within the $h_x <$ 0.5 limit (normalized figure is shown in the inset of Fig. S1(b)). It is noteworthy that the condition $h_x, h_z < 0.5$ was also established in previous studies[1,5] and is applicable to PMA materials from an experimental perspective. Thus, the complete solution of Eq. S(8) takes the following form:

$$sin\theta = sin\theta|_{h_z=0} + fitting\ of\ normalized\ graph\ in\ Fig\ S1(b)\ inset \qquad S(10)$$

We fit this equation with a polynomial of degree n, however, to a good approximation we can neglect second and higher order terms (as $h_x$, $h_z$<<1 since $H_k$>>$H_x$, $H_z$). Incorporating the obtained value of $sin\theta$ and $cos\theta$ in Eq. S(7) with $h_x$, $h_z$<<1, approximation yields the solution:

$$h_{AD}^C = \frac{(h_z - h_x)}{\sqrt{2}} + \frac{1}{2} \qquad S(11)$$

or

$$H_{AD}^C = \frac{1}{\sqrt{2}}(H_z - H_x) + \frac{H_k}{2} \qquad S(12)$$

Equation S(11) or S(12) are the solution for critical anti-damping torque effective field when an additional field $H_Z$ is applied. Further, for $h_z$=0, Eq. S(11) reduces to its previous form as calculated in the literature.[1-4]. Plot of Eq. S(11) for different $h_z$ values in Fig.S1 (c) suggests that $h_z$ modifies the $H_{AD}^C$ values.

Further, from the current and SOT field relation of $H_{AD} = \frac{\hbar}{2e}\frac{j\theta_{SH}}{M_S t_{FM}}$, switching current ($j_c$) corresponding to $H_{AD}^C$ can be obtained as:

$$j_c = \frac{eM_S t_{FM}}{\hbar \theta_{SH}}\left[\sqrt{2}(H_z - H_x) + H_k\right] \qquad S(13)$$

Here, $\theta_{SH}$ is the spin Hall angle of the HM layer and $M_s$ is the saturation magnetization of the heterostructure having a FM layer with thickness $t_{FM}$.

Based on Eq. S(12) and (13), our macrospin analysis suggests that the inclusion of $H_Z$ alters the critical AD-SOT field necessary for switching (Eq. S (12)). As a result, the required critical AD-SOT field increases with applied $H_Z$ values, thereby impacting the critical switching current in a similar manner (Eq. S(13)). Therefore, in the presence of an $H_Z$ field, the current axis is expected to show a shift in magnetization vs. SOT current hysteresis along current axis. The expected shift in the hysteresis loop along the current axis is confirmed by our experimental observations in Fig. 3 of main text.

## S2. Micromagnetic simulations

We performed micromagnetic simulations to examine the magnetization dynamics under the influence of spin-orbit torque and static field $H_Z$. These simulations were performed using MuMax3 micromagnetic solver[6]. We examined a strip with lateral dimensions of 1440 nm X 128 nm which has the similar aspect ratio of Hall bar device used in our experimental study. Further, simulated strip has Pt/Co/Pt material stack parameters, where the ferromagnetic (FM) layer has a thickness of 0.6 nm. The material parameters were sourced from previous experiments[7,8]. Here, saturation magnetization ($M_s$) was $1.4 \times 10^6$ A/m, uniaxial anisotropy constant ($K_u$) calculated from anisotropy field (1.4 T) was $15 \times 10^5$ J/m$^3$, uniform exchange stiffness constant ($A_{ex}$) was set at $16 \times 10^{-12}$ J/m, spin Hall angle ($\theta_{SH}$) was 0.14, interfacial Dzyaloshinskii-Moriya strength (D) was set at 0.05 J/m$^2$ and Gilbert damping parameter ($\alpha$) 0.3 was used. Based on these parameters, a lateral cell size of 2 nm was taken which is less than the exchange length.

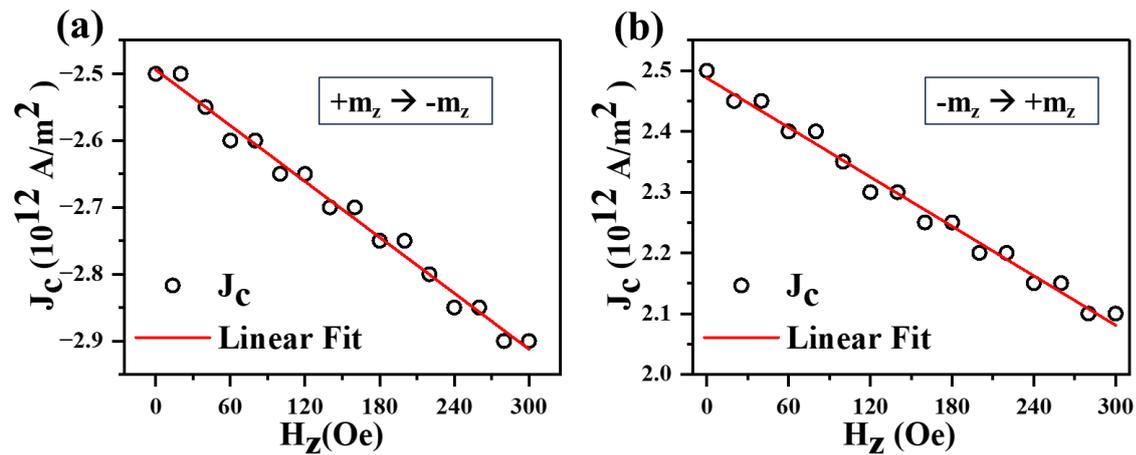

**Fig. S2:** Change of critical switching current as a function of $H_Z$ field when magnetization switches from **(a)** $+m_z$ to $-m_z$, and **(b)** $-m_z$ to $+m_z$

The magnetization evolution was resolved by solving the LLG equation after including the effect of SOT. A 1.85 nS current pulse was applied in presence of a symmetry breaking field Hx ~900 Oe (similar to our experiment) along with a static field along z-direction ($H_Z$). According to the macrospin framework (Eq. S(13)) a linear change in the critical switching current value ($J_c$) was expected with varying $H_Z$. To validate these findings, we performed simulations with different current pulse magnitudes and obtained the critical switching currents. This process was repeated with various $H_Z$ values, and the outcomes are illustrated in Fig. S2 (a) and (b). Here, Fig S2 (a) (Fig. S2 (b)) shows a linear trend of critical current density value required for magnetization switching from $+m_z$ state to $-m_z$ state ($-m_z$ state to $+m_z$ state). The linear trend observed in critical current density change with $H_Z$ corroborates with our macrospin analysis. This correspondence with our macrospin analysis indicates that the modification of the critical SOT field gives rise to this linear trend. Further, we checked the effect of SOT effective field modification on magnetization states. In these simulations, a sufficiently high current pulse, greater than the critical current value, was applied in presence of $H_X$ and various $H_Z$. The magnetization state for each $H_Z$ value was recorded at the time stamp when the state corresponding to $H_Z$= 0 Oe reach to $m_z$= -1 (from +1) after the application of current pulses. The trend of these magnetization states with respect to $H_Z$ is shown in Fig. S2 (c). It is evident from these micromagnetic simulation results that the application of $H_Z$ field alters the magnetic states, indicating the formation of domain states. We have depicted domain magnetization states corresponding to a few $H_Z$ values in Fig. S2 (d).

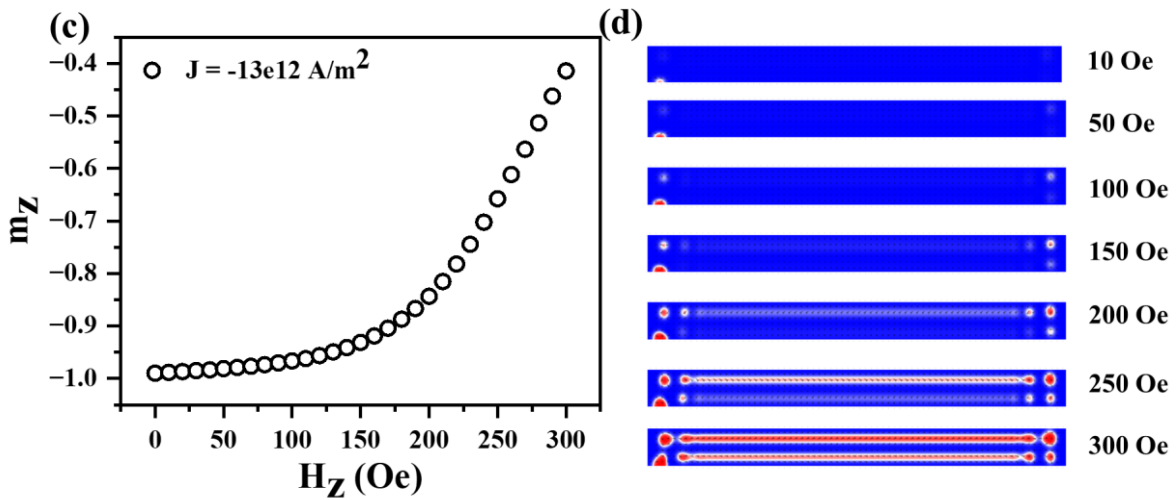

**Fig. S2 (c)** Variation of magnetization ($m_z$) (y-axis) as a consequence of concurrent effect of SOT current ($J_C = -13 \times 10^{12}$ A/m²) and out of plane field ($H_Z$) (x-axis). **(d)** Illustration of magnetization profiles corresponding to some data points from Fig. S2 (c) at different $H_Z$.

Further, to show the universality of the effect we performed similar simulations with a nano-disk with a diameter of 200 nm (results not shown). For various $H_Z$, we found a similar linear trend in the critical current value change along with a non-linear decrease in magnetization state.

## S3. Sample alignment:

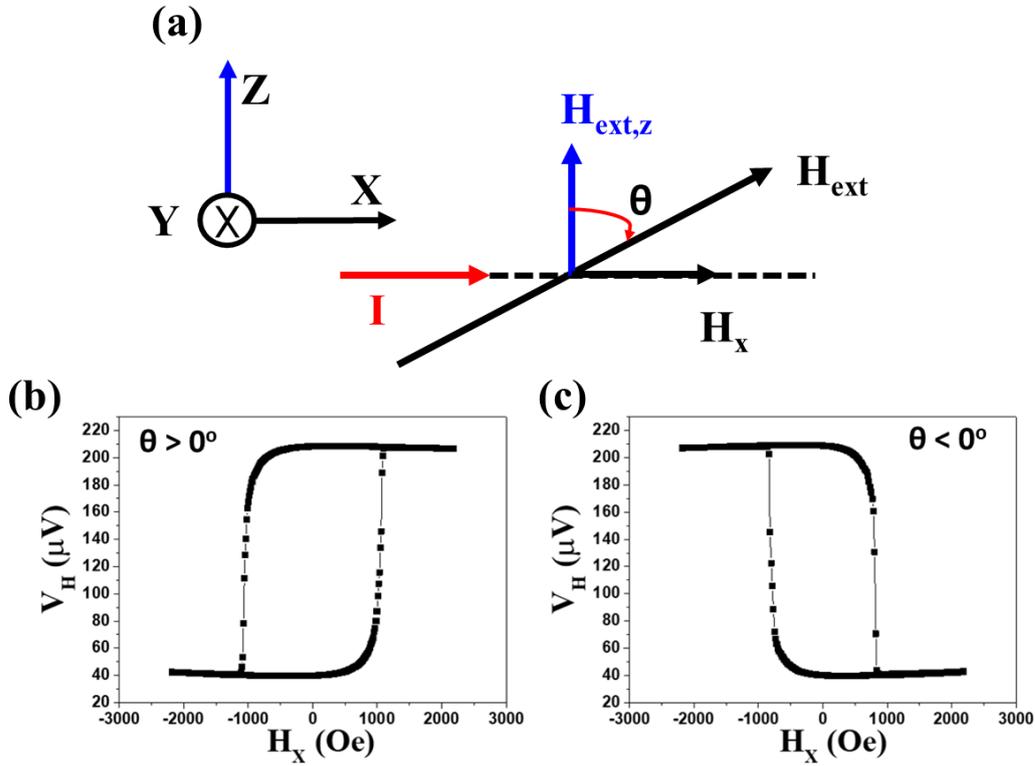

**Fig. S3**. **(a)** Out-of-plane misalignment angle ($\theta$) between applied field ($H_{ext}$) and the current direction producing in-plane and out of plane component of $H_{ext}$. Polarity of hysteresis depending on the component of $H_{ext}$ produced by **(b)** $\theta > 0°$, and **(c)** $\theta < 0°$.

Graphical solution of Eq. S(11) suggest that a small value of out-of-plane field ($H_Z$) is sufficient to alter the values of the critical anti-damping spin-orbit torque (AD-SOT) required for switching. Therefore, before performing any SOT-induced magnetization switching experiment in presence of $H_x$ and $H_Z$, a perfect alignment of sample with $H_x$ is inevitable. During the application of in-plane field along current direction ($H_x$), a small misalignment of applied field ($H_{ext}$) in xz-plane will result as an out-of-plane component of field ($H_{ext,z}$) which will add/subtract with the externally applied $H_Z$ [Note that: $H_x = H_{ext} \sin\theta$ and $H_{ext,z} = H_{ext} \cos\theta$]. Consequently, the additional $H_{ext,z}$, generated as a misaligned component of $H_{ext}$, would misinterpret the effect of externally applied $H_Z$ on SOT-induced magnetization switching. A near perfect alignment of sample with the external field in the device plane is achieved with the help of transport measurements. The misalignment angle ($\theta$) defines the direction (outward/inward) of $H_{ext,z}$ component arising from $H_{ext}$ (Fig. S3a). Therefore, the misalignment angle $\theta>0$ or $\theta<0$ will lead to different polarity of $V_H$ vs. $H_{ext}$ plot (Fig. S3 (b) and (c)). In between these two configurations depicting opposite polarity of the hysteresis curve, $\theta=0$ or a perfect alignment was achieved.

## S4. **Presence of domain states:**

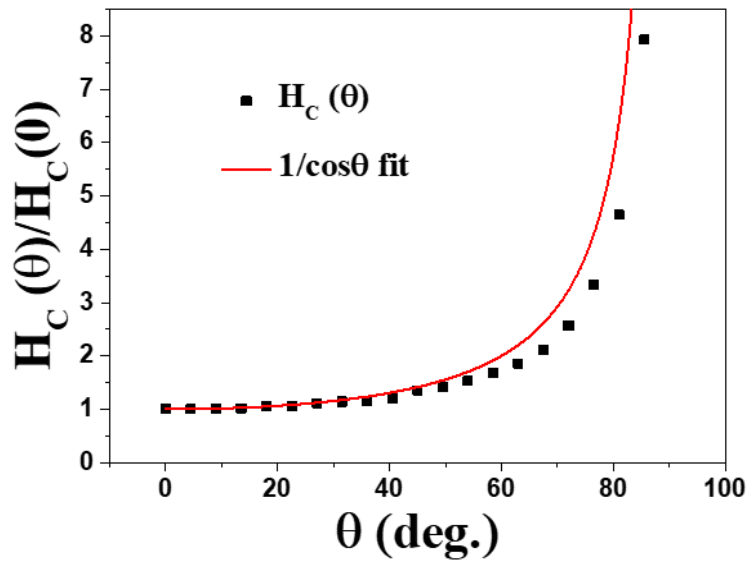

**Fig S4:** Switching field ($H_c$), scaled by value of $H_c$ when the field is applied out of plane to the sample ($H_c(0)$) and plotted as a function of the angle between external field from z-axis (θ) (refer to Fig. S3(a) for coordinates).

In order to understand prevalence of the domain state in the Pt/Co/Pt stack during SOT switching in presence of $H_Z$, we first analyzed the magnetization reversal process in the stack. The two possible reversal mechanisms are: (a) coherent magnetization rotation described by Stoner-Wohlfarth model, and (b) domain propagation mediated magnetization reversal described by Kondorsky model[9]. To probe the reversal mechanism, we obtained the switching field $H_c$ as a function of polar angle θ with respect to the device stack normal. Obtained values of $H_c(θ)$, scaled by switching field at θ=0° ($H_c(0°)$), are plotted in Fig. S4. Here, $H_c(θ)/H_c(0°)$ follows the 1/cosθ dependency as obtained in previous reports[10,11]. This 1/cosθ fit is expected from Kondorsky model. This confirms the presence of domain states in our Pt/Co/Pt system. Furthermore, with symmetry breaking field, a domain state with Néel wall stabilizes in Pt/Co/Pt, we have discussed this in the main text (Fig. 3 and its subsequent discussion).

## S5. Comparison of with-initialization and initialization free states

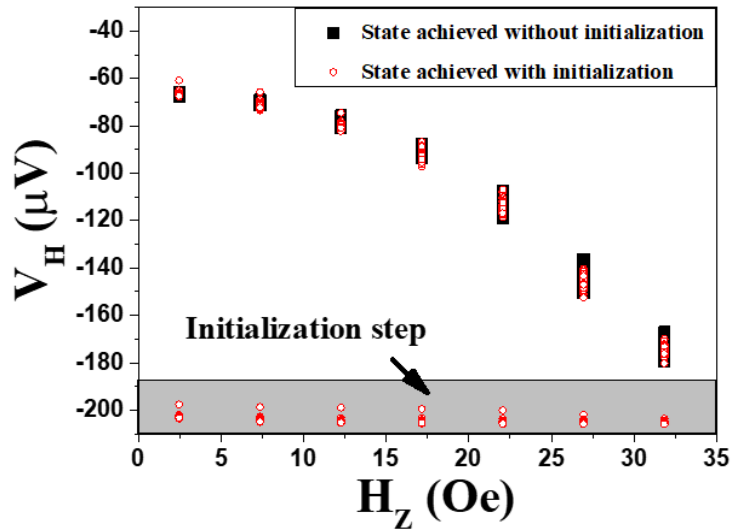
10

**Fig. S5:** Conventional multistate behavior with (red circle) and without (black solid square) initialization step. +60 mA SOT current pulse with different $H_Z$ field stabilizes different states. Initialization step corresponds to the -60 mA pulse

In Figure 4 of the main text, we discuss states with and without an initialization step. To compare whether the multistate levels remain consistent in both cases, we plot Hall voltage ($V_H$) against $H_Z$ field values in Figure S5. This Hall voltage corresponds to the magnetization state, which is achieved by applying a +60 mA current pulse in the presence of various $H_Z$ fields. Figure S5 illustrates the magnetization state in the presence of an initialization current pulse of –60 mA (marked with red circles) and for the initialization-free case (depicted with black solid squares). The initialization state corresponding to the –60 mA pulse, is highlighted with a grey background. The overlapping multistates in Figure S5, with and without initialization, confirm that the states corresponding to various $H_Z$ values remain consistent regardless of whether an initialization step is applied. Additionally, we have repeated the results for 420 pulses to demonstrate the repeatability of the obtained data.

## S6. Exploring the sole effect of $H_Z$ on multistate behavior:

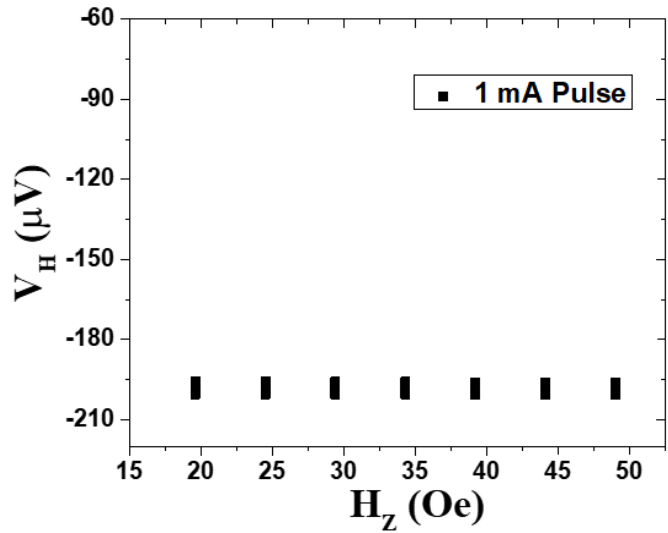

**Fig. S6:** Insufficient current pulse for SOT switching (+1 mA) and varying $H_Z$ showing no change in state ($V_H$).

To demonstrate that the multistate behavior arises from the combined influence of SOT and $H_Z$, rather than solely from $H_Z$, we subjected the system to various $H_Z$ fields and small current pulses of 1mA. It's important to note that a 1 mA pulse is insufficient to induce SOT switching. Therefore, if the multistate behavior persists with small current pulses and different $H_Z$ fields, then it can be attributed to $H_Z$ alone. However, $H_Z$ field alone did not exhibit any multistate behavior, and the states remained at a single level. These results are presented in Figure S6.

Please note that in Figure S5, the multistate behavior arises when the applied current exceeds the critical current value (applied current (60mA) > switching currents (~40 mA)). In Figure S5, the applied current generates a sufficient SOT effective field to switch the magnetization and exhibit multistate behavior when integrated with the $H_Z$ field. In comparison, Figure S6 maintains a fixed state for various values of $H_Z$ while the SOT current is only 1 mA. These results confirm that the $H_Z$ field alone cannot induce multistate behavior and that the combined effect of both SOT and $H_Z$ is necessary to achieve this effect.

## S7. <u>Retention of states after switching off the field $H_Z$:</u>

The $H_Z$ field-induced multistate behavior holds promising applications in analog computing due to its high bit density per cell. Magnetic devices with a multistate behavior are of great potential due to their non-volatility. To assess the stability of these multistates, we examined the states after turning off the magnetic field $H_Z$. In these experiments, initially, a 'State 1' with $H_Z$=0 Oe is achieved (achieved by setting $H_Z$ to 0 Oe and applying a current of +60 mA). This state exhibits the highest magnitude among all multistates, as depicted in Figure 2 of the main text. Transitioning from 'State 1' to a different state requires the application of a 60 mA pulse in the presence of a field $H_Z \neq 0$ Oe. Once the new state is reached, we confirm its stability by turning off the $H_Z$ field and measuring the Hall voltage: (a) immediately after switching off the field and (b) after a defined time interval. The results are presented in a bar graph shown in Figure S7. Here, the y-axis represents Hall voltage, thus distinguishing between multiple states, while the x-axis denotes the current pulse, $H_Z$ field, and time interval conditions. For a better understanding, the conditions for each bar line are also described in a table below.

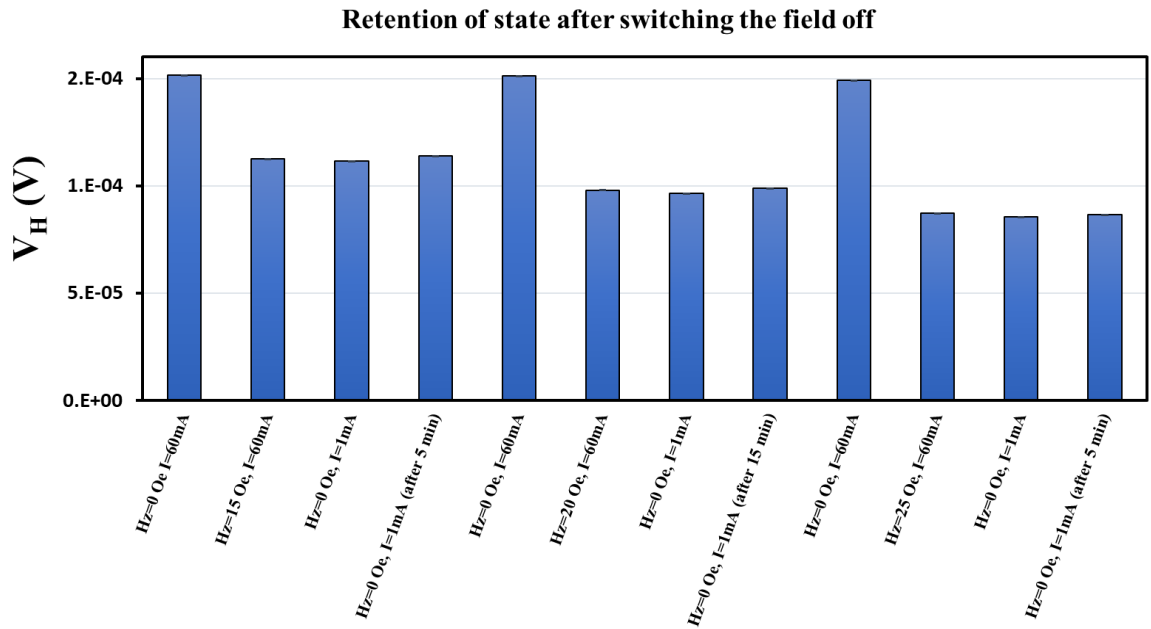

**Fig. S7** Retention of different states after switching off the $H_Z$ field. States are read after various time intervals.

| S.No. | Write current | Read current | $H_Z$ field | Time condition |
|---|---|---|---|---|
| 1. | 60 mA | 1 mA | 0 Oe | - |
| 2. | 60 mA | 1 mA | 15 Oe | - |
| 3. | - | 1 mA | 0 Oe | Read immediately after $H_Z$=0 Oe |
| 4. | - | 1 mA | 0 Oe | Read after 5 min. |
| 5. | 60 mA | 1 mA | 0 Oe | - |
| 6. | 60 mA | 1 mA | 20 Oe | - |
| 7. | - | 1 mA | 0 Oe | Read immediately after $H_Z = 0$ Oe |
| 8. | - | 1 mA | 0 Oe | Read after 15 min. |
| 9. | 60 mA | 1 mA | 0 Oe | - |
| 10. | 60 mA | 1 mA | 25 Oe | - |
| 11. | - | 1 mA | 0 Oe | Read immediately after $H_Z = 0$ Oe |
| 12. | - | 1 mA | 0 Oe | Read after 5 min. |

These results suggest that after achieving a state by the combine effect of $H_z$ and SOT when $H_z$ is turned off, magnetization retains at this state for a long duration. Therefore, a continuous applied field ($H_z$) is not required to sustain a magnetic state, making this $H_z$ field-induced multistate behavior an energy-efficient approach.

## S8. <u>Multistate behavior in W/Pt/Co/AlOx PMA stack:</u>

To demonstrate the universality of multistate occurrence in perpendicularly magnetized stacks, we conducted experiments using a Ta(3)/W(0.85)/Pt(3)/Co(0.6)/AlOx(1.3) PMA stack, where the thicknesses of each layer are indicated in parentheses. We induced SOT-driven switching in the presence of an $H_Z$ field to investigate multistate behavior. Remarkably, the application of $H_Z$ during SOT switching led to multistate behavior, as illustrated in Figure S8 (a).

Interestingly, we observed a trend of reduction in magnetization states in the W/Pt/Co/AlOx stack, similar to the trend observed in the Pt/Co/Pt stack discussed in the main text. Figure S8 (b) depicts this reduction trend under various $H_Z$ fields (x-axis) in presence of a SOT current pulses (~45 mA). It's important to note that after aligning the W/Pt/Co/AlOx devices with an in-plane field (using the method described in section S3), both Figure S8 (a) and S8 (b) were obtained in the presence of a symmetry-breaking field $H_x$= 900 Oe. This universal trend of magnetization state reduction confirms the general applicability of the proposed method across different material configurations.

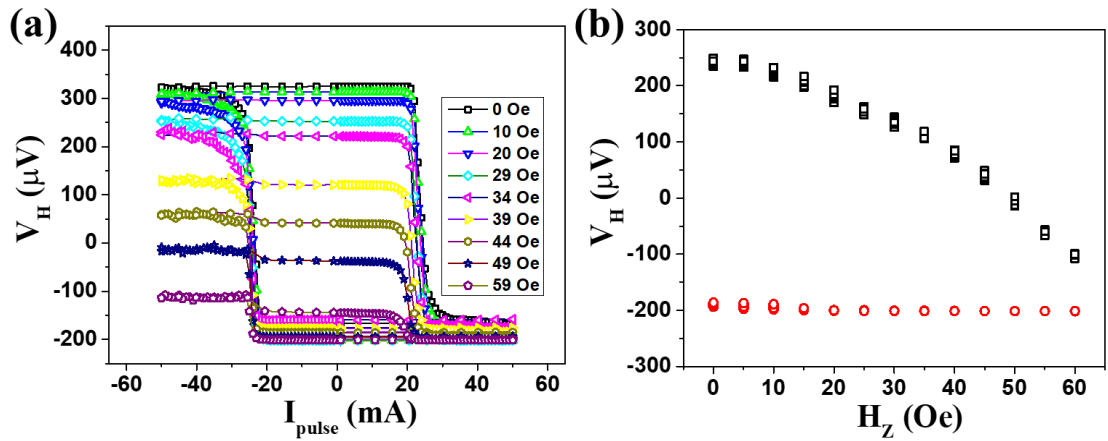

**Fig S8 (a)** SOT switching hysteresis after the application of various $H_Z$ fields in W/Pt/Co/AlO$_x$ stack. For each $H_Z$ field value Hall voltage (y-axis) is recorded after an application of 5 ms wide current pulses (x-axis). **(b)** Trend of the reduction of magnetization states (i.e., Hall voltage at saturation, here it is read after the application of 5ms current pulse of +45 mA) as a function of $H_Z$.

# References:

[1] K.-S. Lee, S.-W. Lee, B.-C. Min, and K.-J. Lee, Threshold current for switching of a perpendicular magnetic layer induced by spin Hall effect, Applied Physics Letters **102**, 112410 (2013).

[2] T. Taniguchi, Theoretical condition for switching the magnetization in a perpendicularly magnetized ferromagnet via the spin Hall effect, Physical Review B **100**, 174419 (2019).

[3] D. Zhu and W. Zhao, Threshold Current Density for Perpendicular Magnetization Switching Through Spin-Orbit Torque, Physical Review Applied **13**, 044078 (2020).

[4] T. Taniguchi, S. Mitani, and M. Hayashi, Critical current destabilizing perpendicular magnetization by the spin Hall effect, Physical Review B **92**, 024428 (2015).

[5] L. Liu, O. J. Lee, T. J. Gudmundsen, D. C. Ralph, and R. A. Buhrman, Current-Induced Switching of Perpendicularly Magnetized Magnetic Layers Using Spin Torque from the Spin Hall Effect, Physical Review Letters **109**, 096602 (2012).

[6] A. Vansteenkiste, J. Leliaert, M. Dvornik, M. Helsen, F. Garcia-Sanchez, and B. Van Waeyenberge, The design and verification of MuMax3, AIP Advances **4**, 107133 (2014).

[7] R. Posti, A. Kumar, D. Tiwari, and D. Roy, Emergence of considerable thermoelectric effect due to the addition of an underlayer in Pt/Co/Pt stack and its application in detecting field free magnetization switching, Applied Physics Letters **121**, 223502 (2022).

[8] P. P. J. Haazen, E. Murè, J. H. Franken, R. Lavrijsen, H. J. M. Swagten, and B. Koopmans, Domain wall depinning governed by the spin Hall effect, Nature Materials **12**, 299 (2013).

[9] E. J. J. P. Kondorsky, On hysteresis in ferromagnetics, **2**, 161 (1940).

[10] S. Kim, P.-H. Jang, D.-H. Kim, M. Ishibashi, T. Taniguchi, T. Moriyama, K.-J. Kim, K.-J. Lee, and T. Ono, Magnetic droplet nucleation with a homochiral N\'eel domain wall, Physical Review B **95**, 220402 (2017).

[11] O. J. Lee, L. Q. Liu, C. F. Pai, Y. Li, H. W. Tseng, P. G. Gowtham, J. P. Park, D. C. Ralph, and R. A. Buhrman, Central role of domain wall depinning for perpendicular magnetization switching driven by spin torque from the spin Hall effect, Physical Review B **89**, 024418 (2014).